\documentclass[aps,nofootinbib,floatfix,amsmath,amssymb,superscriptaddress,preprint]{revtex4-1}
\pdfoutput=1

\usepackage{preamble}

\newcommand{\modu}[1]{\left|{#1}\right|}
\newcommand{\deq}{\coloneqq}

\newcommand{\eff}{\epsilon}
\newcommand{\purity}{\rho}

\begin{document}

\title{Testing quark mixing in minimal left-right symmetric models with $b$-tags at the LHC}
\author{Andrew Fowlie}
\email{Andrew.Fowlie@KBFI.ee}
\affiliation{National Institute of Chemical Physics and Biophysics, Ravala 10,
Tallinn 10143, Estonia}

\author{Luca Marzola}
\email{Luca.Marzola@ut.ee}
\affiliation{Laboratory of Theoretical Physics, Institute of Physics, University of Tartu, Ravila 14c, Tartu 50411, Estonia}

\date{\today}

%%%%%%%%%%%%%%%%%%%%%%%%%%%%%%%%%%%%%%%%%%%%%%%%%%%%%%%%%%%%%%%%%%%%%%%%%%%%%%%%

\begin{abstract}
Motivated by a hint in a CMS search for right-handed $W$-bosons in $eejj$ final states, we propose an experimental test of quark-mixing matrices in a general left-right symmetric model, based on counting the numbers of $b$-tags from right-handed $W$-boson hadronic decays. 
We find that, with our test, differences between left- and right-handed quark-mixing matrices could be detected at the LHC with \roots{14}.
With an integrated luminosity of about $20\invfb$, our test is sensitive to right-handed quark-mixing angles as small as about $30\degree$ and with $3000\invfb$, our test's sensitivity improves to right-handed mixing angles as small as about $7.5\degree$. 
Our test's sensitivity might be further enhanced by tuning $b$-tagging efficiency against purity.

\end{abstract}
%%%%%%%%%%%%%%%%%%%%%%%%%%%%%%%%%%%%%%%%%%%%%%%%%%%%%%%%%%%%%%%%%%%%%%%%%%%%%%%%

\maketitle

\section{Introduction}
An unexplained feature of the Standard Model (SM)\cite{Glashow:1961tr,Salam:1968rm,Weinberg:1967tq} is that left-right symmetry is broken; only left-handed fermions take part in weak interactions\cite{Lee:1956qn}. In the 1970s, Georgi and Glashow\cite{Georgi:1974sy}, amongst others\cite{Pati:1973uk,Pati:1974yy,Fritzsch:1974nn,Gursey:1975ki,Buras:1977yy}, realized that puzzling aspects of the SM could be explained if, at high energy, nature is symmetric under a simple or semi-simple Lie group. This popular proposal became known as a grand unified theory (GUT) and provided an ideal framework in which to restore left-right symmetry at high energy\cite{Pati:1973uk,Pati:1974yy,Mohapatra:1974hk,Mohapatra:1974gc,Senjanovic:1975rk,Beg:1977ti,Senjanovic:1978ev}. 

A left-right symmetric GUT gauge group can be spontaneously broken to the SM gauge group via a left-right symmetric product gauge group.
%; this is the case in the Pati-Salam model\cite{Pati:1973uk,Pati:1974yy}. 
Minimal realizations of the latter contain the product gauge group  $\text{SU}(2)_L \times \text{SU}(2)_R$, as well as a discrete symmetry that ensures that the representations and couplings for the $\text{SU}(2)_L$ and $\text{SU}(2)_R$ gauge groups are indeed left-right symmetric. Generalized parity, $\mathcal{P}$, and generalized charge conjugation, $\mathcal{C}$, are common candidates for that discrete symmetry. At low energy, the minimal left-right symmetric gauge group is broken to the familiar SM gauge group and neither $\mathcal{P}$ nor $\mathcal{C}$ symmetry is preserved.

After the spontaneous symmetry breaking of both $\text{SU}(2)_L$ and $\text{SU}(2)_R$, a left-right symmetric model includes massive $W_{R,L}$-bosons,\footnote{The label on a gauge boson refers to the handedness of the fermions with which it interacts.} massive quarks and two distinct quark-mixing matrices \see{Maiezza:2010ic}. These result from the
misalignment between the quark mass eigenstates and the $\text{SU}(2)_R$ or $\text{SU}(2)_L$ interaction eigenstates, with the SM CKM matrix\cite{Kobayashi:1973fv,Cabibbo:1963yz} (henceforth LH CKM matrix) describing the resulting quark mixing in the latter case. 
% , resulting from the misalignment between the quark mass eigenstates and the left-handed quark $\text{SU}(2)_L$ interaction eigenstates, and
% \item a right-handed (RH) quark-mixing matrix, resulting from the misalignment between the quark mass eigenstates and the right-handed quark $\text{SU}(2)_R$ interaction eigenstates.
% \end{itemize}
It was recently shown in \refcite{Senjanovic:2014pva}, following earlier work in \refcite{Kiers:2002cz,Zhang:2007fn}, that in minimal left-right symmetric GUTs, the LH CKM matrix and the RH mixing matrix are approximately identical, modulo complex phases. In a general left-right symmetric GUT, this is possible, though not compulsory.

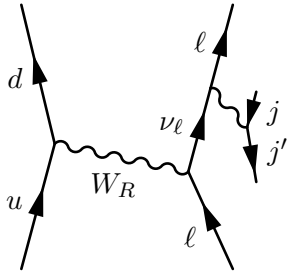
\begin{figure}[h]
\begin{feynman}{production}
\fmfleft{i1,i2}
\fmfright{o1,o2,o3,o4}
\fmf{fermion}{i1,v1,i2}
\fmf{phantom,label=$u$,label.side=left}{i1,v1}
\fmf{phantom,label=$d$,label.side=left}{v1,i2}
\fmf{boson,label=\rw}{v1,v2}
\fmf{fermion}{o1,v2,v3,o4}
\fmf{phantom,label=$\ell$}{o1,v2}
\fmf{phantom,label=$\ell$}{o4,v3}
\fmf{phantom,label=$\nu_\ell^R$,label.side=left}{v2,v3}
\fmffreeze
\fmf{boson,label=\rw}{v3,v4}
\fmf{fermion}{o3,v4,o2}
\fmf{phantom,label=$j$,label.side=left}{o3,v4}
\fmf{phantom,label=$j^\prime$,label.side=right}{o2,v4}
\end{feynman}
\caption{Feynman diagram for the production and decay of a right-handed $W$-boson \rw at the LHC.}
\label{Fig:Feyn}
\end{figure}

Left-right symmetric models are nowadays particularly interesting in light of an experimental hint from the Large Hadron Collider (LHC). In a recent CMS analysis\cite{Khachatryan:2014dka}, the number of events presenting two electrons (with no charge requirement imposed, \ie $e^-e^+$, $e^+ e^+$ or $e^- e^-$) and two jets in the final state exceeded the prediction of the SM. 
% Once events were binned with respect to a kinematic variable, the excess occurred in a single bin. A discrepancy that large and in that particular bin would occur with a probability of $0.0050$, were the SM true. The analysis, however, searched for discrepancies in $20$ bins. The probability that a discrepancy that large occurred in \emph{any} of the $20$ bins might be appreciable. 
Whilst the excess might have a mundane explanation, such as a statistical fluctuation or a systematic error, we regard it as an intriguing hint. In fact, the detected anomaly could be explained by the production and subsequent decay of a right-handed $W$-boson in a left-right symmetric model (\reffig{Fig:Feyn}),
\beq
\label{Eq:DecayChain}
q\bar q^\prime \to \rw \to e \nu_e^R  \to e e \rw \to e e j j,  
\eeq
provided the right-handed $W$-boson has a mass of about $2\tev$ and only the right-handed electron-neutrino is lighter than about $2\tev$\cite{Heikinheimo:2014tba,Deppisch:2014qpa,Aguilar-Saavedra:2014ola}. 

%%%%%%%%%%%%%%%%%%%%%%%%%%%%%%%%%%%%%%%%%%%%%%%%%%%%%%%%%%%%%%%%%%%%%%%%%%%%%%%%
%%%%%%%%%%%%%%%%%%%%%%%%%%%%%%%%%%%%%%%%%%%%%%%%%%%%%%%%%%%%%%%%%%%%%%%%%%%%%%%%

With the recent experimental hint in mind\cite{Khachatryan:2014dka}, 
% though without much loss of generality, 
we consider a scenario in which there exists a heavy right-handed $W$-boson and show that the equality of the LH CKM matrix and the RH mixing matrix could be tested at the LHC. As shown below, our method categorizes the hadronic decays of the new gauge boson by their number of $b$-tags\footnote{The mass of the bottom quark is such that it travels within the LHC detectors before decaying at a displaced vertex to highly energetic jets. From these features, $b$-jets can be identified or \ic{tagged} by $b$-tagging algorithms \see{Chatrchyan:2012jua}. Because top quarks decay into bottom quarks, top quarks result in a $b$-jet which can be $b$-tagged.} and quantifies the probability of obtaining the same result under the assumption that the RH mixing matrix matches the LH CKM one. The proposed procedure is therefore able to quantify the discrepancy between the quark mixings of the two chiral sectors in a model-independent way and constitutes a new collider test of minimal left-right models that complements the model-dependent results brought by meson-oscillation experiments\cite{Beall:1981ze,PhysRevD.40.1569} and low-energy observables \see{Barenboim:1996nd,Blanke:2011ry}.

%%%%%%%%%%%%%%%%%%%%%%%%%%%%%%%%%%%%%%%%%%%%%%%%%%%%%%%%%%%%%%%%%%%%%%%%%%%%%%%%
\section{Methodology}\label{sec:Method-LR}
%%%%%%%%%%%%%%%%%%%%%%%%%%%%%%%%%%%%%%%%%%%%%%%%%%%%%%%%%%%%%%%%%%%%%%%%%%%%%%%%
The RH mixing matrix affects the rate at which right-handed $W$-bosons are produced from two protons and the right-handed $W$-boson's branching fractions to quarks in the decay chain in \refeq{Eq:DecayChain}. The production cross section depends on three unknown quantities: the RH mixing matrix, the right-handed gauge coupling at low energy, $g_R(\mw)$, and the right-handed $W$-boson mass.\footnote{The right-handed gauge coupling at low energy, $g_R(\mw)$, might differ from $g_L(\mw)$ by renormalization group running, even if they are equal at a high energy.} The branching fractions in the final hadronic decay, however, depend on only the RH mixing matrix. Thus, to investigate the RH mixing matrix, the right-handed $W$-boson's hadronic decay is the best place to start. 

We parameterize the RH mixing matrix in the standard way\cite{Chau:1984fp,Beringer:1900zz}, \ie as the product of rotations on three planes in the basis of the quark fields $(d,s,b)^T$;
\beq
V_R = 
\left( 
\begin{array}{ccc}
1 & 0 & 0 \\
0 & c_{23} & s_{23} \\
0 & -s_{23} & c_{23} 
\end{array} 
\right)
\left( 
\begin{array}{ccc}
c_{13} & 0 & s_{13} \\
0 & 1 & 0 \\
-s_{13} & 0 & c_{13} 
\end{array} 
\right)
\left( 
\begin{array}{ccc}
c_{12} & s_{12} & 0 \\
-s_{12} & c_{12} & 0 \\
0 & 0 & 1 
\end{array} 
\right),
\eeq
where $c_{ij} \equiv \cos\theta_{ij}$ and $s_{ij} \equiv \sin \theta_{ij}$ with a superscript `$R$' left understood.  
In principle, the RH mixing matrix contains six physical phases on top of three mixing angles. However, as will become soon apparent, our work is not sensitive to these quantities and therefore we chose to disregard them for the sake of simplicity. 

Given the above mixing matrix, we calculate the right-handed $W$-boson's hadronic branching fractions through the Feynman rule
\begin{align}
  \Diagram{
  & & & & \hspace{.2 cm} \vertexlabel{_}{\displaystyle \bar{q}_R}\\
  \vertexlabel{^}{\displaystyle \rw} \hspace{.3 cm}& g & fuV & &\\
  & & fdA & & \\
  & & & & \hspace{.2 cm} \vertexlabel{^}{\displaystyle q_R^\prime}
  }
  \quad 
  =
  \;
  ig_R V^R_{qq^\prime} \gamma^\mu,
\end{align}
and assume that the right-handed $W$-boson is much heavier than the top quark, $M_{\rw}\approx 2\tev \gg m_t$, such that all quark masses are negligible. Motivated by the experimental hint\cite{Khachatryan:2014dka}, we also assume that the right-handed muon-neutrino is heavier than the right-handed $W$-boson, $m_{\nu^R_\mu} > M_{\rw}$, but that $m_{\nu^R_e} < M_{\rw}$. On top of that we neglect $W_L$-$W_R$ mixing.

Although the right-handed $W$-boson's hadronic branching fractions can be straightforwardly computed, the $b$-tagging algorithms adopted in an experimental analysis are still imperfect:
\begin{itemize}
\item The efficiency, $\eff$, is the probability that a genuine $b$-jet is $b$-tagged. With an appreciable probability, $1-\eff$, a genuine $b$-jet might not be $b$-tagged. We assume that $\eff=0.7$.
\item The purity, $\purity$, is the probability that a genuine light-jet is not $b$-tagged. With a small probability, $1-\purity$, a genuine light jet might be $b$-tagged.\footnote{We refer to first- and second-generation quarks as light quarks.} We assume that $\purity=0.99$.
\end{itemize}
Thus, \emph{any} right-handed $W$-boson hadronic decay could actually result in $0$, $1$ or $2$ $b$-tags, even if the right-handed $W$-boson decayed to only light quarks.

By combining our knowledge of the imperfections of $b$-tagging with the tree-level right-handed $W$-boson hadronic branching fractions, we then expect that right-handed $W$-bosons that decay hadronically result in $0$, $1$ or $2$ $b$-tags with the following probabilities,
\begin{align}
\label{Eq:Prob_btags_0}
p_0 & \equiv p(\text{$0$ $b$-tags from \rw hadronic decay}) \propto
  \purity^2 C_0 + \purity \(1-\eff\) C_1 + \(1-\eff\)^2 C_2 
  ,\\
\label{Eq:Prob_btags_1}
p_1 & \propto
  2 \purity \(1-\purity\) C_0 + \purity \eff C_1 + \(1-\purity\)\(1-\eff\) C_1 + 2 \eff \(1-\eff\) C_2 
  ,\\
p_2 &\propto
\label{Eq:Prob_btags_2}
  \(1-\purity\)^2 C_0 + \eff\(1-\purity\) C_1 + \eff^2 C_2 
  ,
\end{align}
which include all possible mistaggings with the appropriate weights. In the above expressions we omitted a normalization constant and defined
\begin{align} 
   \begin{split}\label{eq:def_C0}    
       C_{0} 
      \deq & 
      \modu{V^R_{11}}^2 + \modu{V^R_{12}}^2 
      + 
      \modu{V^R_{21}}^2 + \modu{V^R_{22}}^2
      \\      
      = &
     1 + \cos^2\theta_{13} \cos^2\theta_{23},
	 \end{split}\\
   \begin{split}\label{eq:def_C1}
      C_{1} 
      \deq &
      \modu{V^R_{31}}^2 + \modu{V^R_{32}}^2 
      + \modu{V^R_{23}}^2 + \modu{V^R_{13}}^2
      \\      
      = &
      2 \(1 - \cos^2\theta_{13}\cos^2\theta_{23}\),
	  \end{split}\\
   \begin{split}\label{eq:def_C2}
      C_{2} 
      \deq & 
      \modu{V^R_{33}}^2 = \cos^2\theta_{13} \cos^2\theta_{23}.
      \end{split}
\end{align} 
As anticipated, the probabilities are independent of phases in the RH mixing matrix and, interestingly, depend on only the $\theta_{13}$ and $\theta_{23}$ mixing angles of the former. The dependence on the remaining mixing angle, $\theta_{12}$, is lost because this quantity regulates the mixing of first- and second-generation light quarks and therefore cannot affect the expected fraction of $b$-jets or light jets.

We assume that the production cross section for the right-handed $W$-boson is such that we expect that $10$ of the $14$ events in the $\sim2\tev$ bin in \refcite{Khachatryan:2014dka} result from the decay chain in \refeq{Eq:DecayChain}. This can be achieved by tuning the right-handed $W$-boson coupling and mass. We expect that the remaining $4$ events result from SM backgrounds, as indicated in \refcite{Khachatryan:2014dka}. However, with so few events, it is impossible to infer interesting information about the RH mixing matrix. 
Thus, we refer to a $\roots{14}$ scenario with an integrated luminosity identical to that in \refcite{Khachatryan:2014dka}, $\sim20\invfb$ and scale the numbers of signal and background events in \refcite{Khachatryan:2014dka} by the ratio of the corresponding cross sections at \roots{14} and \roots{8}. Consequently, at \roots{14} we expect $s=67.0$ signal events\cite{Kirsanov:2012} and $b=15.6$ background events\cite{Czakon:2013goa} and the increased number of signal events makes it possible to study the RH mixing matrix in this scenario.
% More in detail, 
With our Eq.~\ref{Eq:Prob_btags_0}, \ref{Eq:Prob_btags_1} and \ref{Eq:Prob_btags_2}, we will show that, if our alternative hypothesis is correct, future LHC experiments would have the power to reject the null hypothesis that $V_L=V_R$ with at least $95\%$ confidence by counting the numbers of $b$-tags.

For this purpose we consider two cases, which we regard as hypotheses in the statistical test performed below:
\begin{itemize}
\item The null hypothesis, $H_0$: the RH mixing matrix is equal to the LH CKM matrix,
\beq
V_R = V_L
\eeq
with $V_L$ fixed by the usual SM quark mixing \see{Giunti:2007ry}.
\item The alternative hypothesis, $H_1$: the RH mixing matrix is independent of the LH CKM matrix. 
\end{itemize}

From our Eq.~\ref{Eq:Prob_btags_0}, \ref{Eq:Prob_btags_1} and \ref{Eq:Prob_btags_2}, we then calculate the expected numbers of events with $i$ $b$-tags from a right-handed $W$-boson hadronic decay, 
\beq
s_i = p_i \times s,
\eeq
where $s$ is the total number of expected signal events. Unfortunately, the signal region is contaminated with SM background events. The dominant SM background is $t\bar t$. We make the approximation that all SM backgrounds result in the same $b$-tag distribution as that of $t\bar t$ production, such that
\beq
b_i = p(i \text{ $b$-tags from $t\bar t$}) \times b,
\eeq
where $b$ is the total number of expected background events and the probability is calculated in a manner analogous to that in Eq.~\ref{Eq:Prob_btags_0}, \ref{Eq:Prob_btags_1} and \ref{Eq:Prob_btags_2}. This approximation is conservative, because  $t\bar t$ contaminates all $b$-tag categories with appreciable probabilities.

In a counting experiment, such as which we propose, the numbers of observed events in each $b$-tag category, $o_i$, are Poisson distributed,
\beq
\label{Eq:Poisson}
o_i \sim \text{Po}(s_i+b_i)
\eeq
and independent of each other.

Throughout the following discussion, our notation is such that if a quantity is calculated under
the null hypothesis, it is superscripted with a zero, whereas if it is calculated under the alternative hypothesis, it is superscripted with a one. 
Our methodology is that for given mixing angles in the RH mixing matrix:
\begin{enumerate}
\item From the Poisson distributions in \refeq{Eq:Poisson}, we sample $1000$ Monte-Carlo (MC) measurements of the numbers of observed events in each $b$-tag category, $o_i^1$, with the alternative hypothesis. The number of signal events in each $b$-tag category is a function of the RH mixing angles. %, $s_i^1 = f(\theta)$. 

\item For each of the $1000$ MC measurements, we calculate a log-likelihood ratio test-statistic (LLR) associated with the null hypothesis that $V_L = V_R$ and the alternative hypothesis;
\begin{align}
\label{Eq:LLR0}
\text{LLR} &= -2\ln \frac{\mathcal{L}(o_i^1\,|\,H_0)}{\max\limits \mathcal{L}(o_i^1\,|\,H_1)} \\
\label{Eq:LLR}
&= -2\sum_i \ln \frac{(s_i^0 + b_i)^{o_i^1}e^{-(s_i^0 + b_i)}}{o_i^1!} + 2\sum_i \ln \max \frac{(s_i^1 + b_i)^{o_i^1}e^{-(s_i^1 + b_i)}}{o_i^1!},
\end{align}
where $\mathcal{L}$ are likelihood functions. A likelihood function $\mathcal{L}(d \,|\,H)$ returns the probability of observing the data $d$ from an experiment, within the framework specified by the hypothesis $H$. Thus in \refeq{Eq:LLR0}, $\mathcal{L}(o_i^1\,|\,H_0)$ returns the probability of obtaining the set $o_i^1$ for the observed numbers of events in each $b$-tag category, under the assumption of our null hypothesis $H_0$. The resulting probability is compared by means of the LLR with the corresponding probability obtained under the alternative hypothesis, $\max\limits \mathcal{L}(o_i^1\,|\,H_1)$, where the likelihood is maximized by tuning the RH mixing matrix elements. 

For each hypothesis, the likelihood function is a Poisson distribution. The hypotheses $H_0$ and $H_1$, however, specify the expected numbers of signal events: $s_i^0$ in the null hypothesis and $s_i^1$ in the alternative hypothesis, in the first and second term of \refeq{Eq:LLR} respectively.

By Wilks' theorem, because the expected numbers of events, $s_i^0+b_i$, are greater than about $5$, in the null hypothesis the LLR is approximately $\chi^2$-distributed with $3$ degrees of freedom,\footnote{There are $3$ approximately Gaussian contributions to the likelihood. In the first term in \refeq{Eq:LLR}, no parameters are tuned, resulting in $3$ degrees of freedom. In the second term, $18$ parameters in the RH mixing matrix are tuned, resulting in $0$ degrees of freedom. Thus, there are $3-0=3$ degrees of freedom in the LLR.}
\beq
\text{LLR} \sim \chi^2_3.
\eeq
The $p$-value is the probability of obtaining such a large test-statistic by chance, were the null hypothesis true. 

\item Finally, we find the median and $68\%$ confidence interval for the $p$-value, by considering all of our MC experiments. Our ordering rule for the $68\%$ confidence interval is that $16\%$ of our MC experiments resulted in $p$-values above the interval and that $16\%$ resulted in $p$-values below the interval.
\end{enumerate}
 
%%%%%%%%%%%%%%%%%%%%%%%%%%%%%%%%%%%%%%%%%%%%%%%%%%%%%%%%%%%%%%%%%%%%%%%%%%%%%%%%
\section{Results}
\begin{figure}[t]
\centering
\subfloat[][$(\theta_{23},\theta_{13})$ plane.]{\label{fig:2d}
\includegraphics[width=.48\textwidth,valign=t]{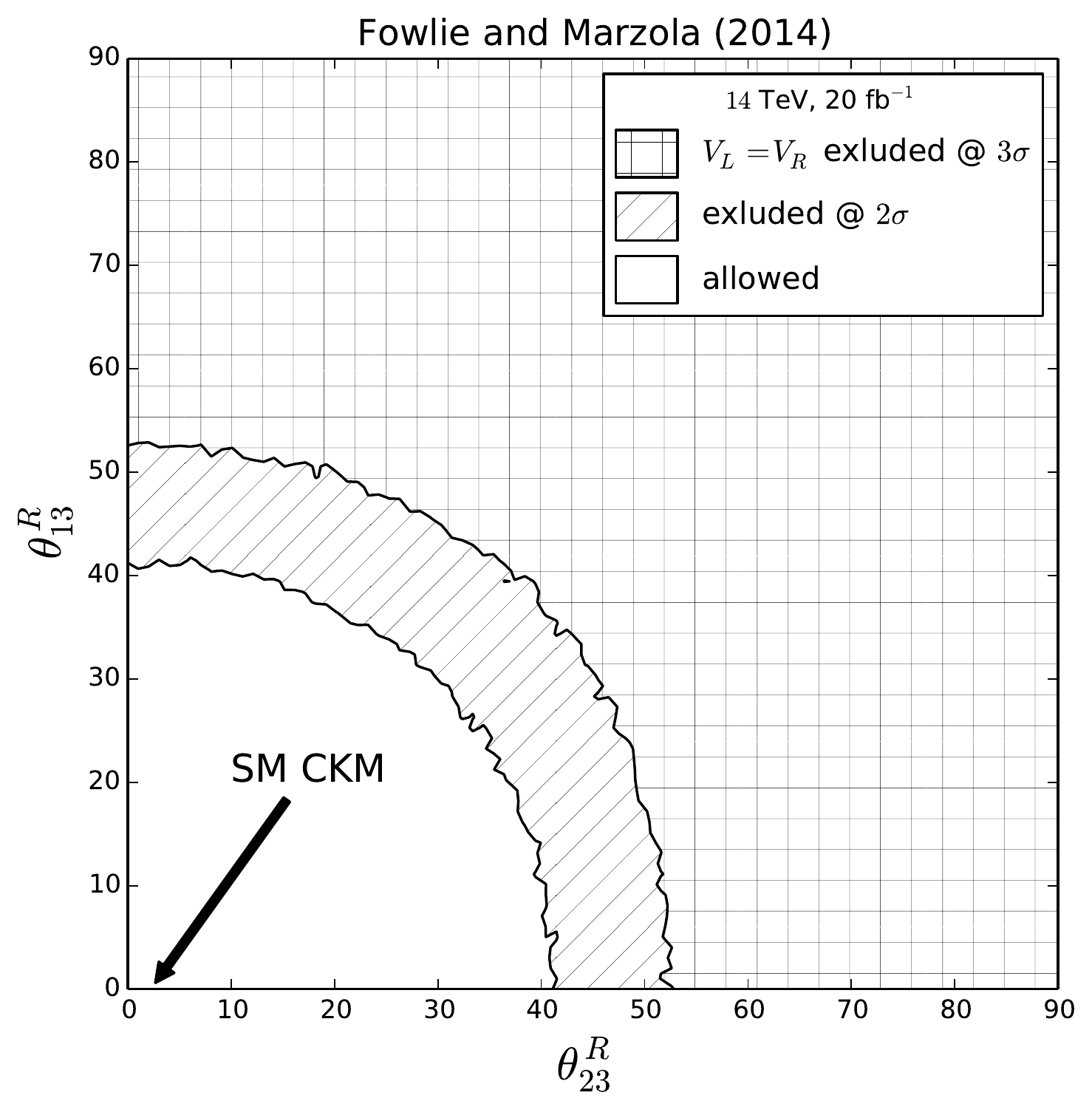}
}
\subfloat[][Universal mixing angle.]{\label{fig:1d}
\includegraphics[width=.48\textwidth,valign=t]{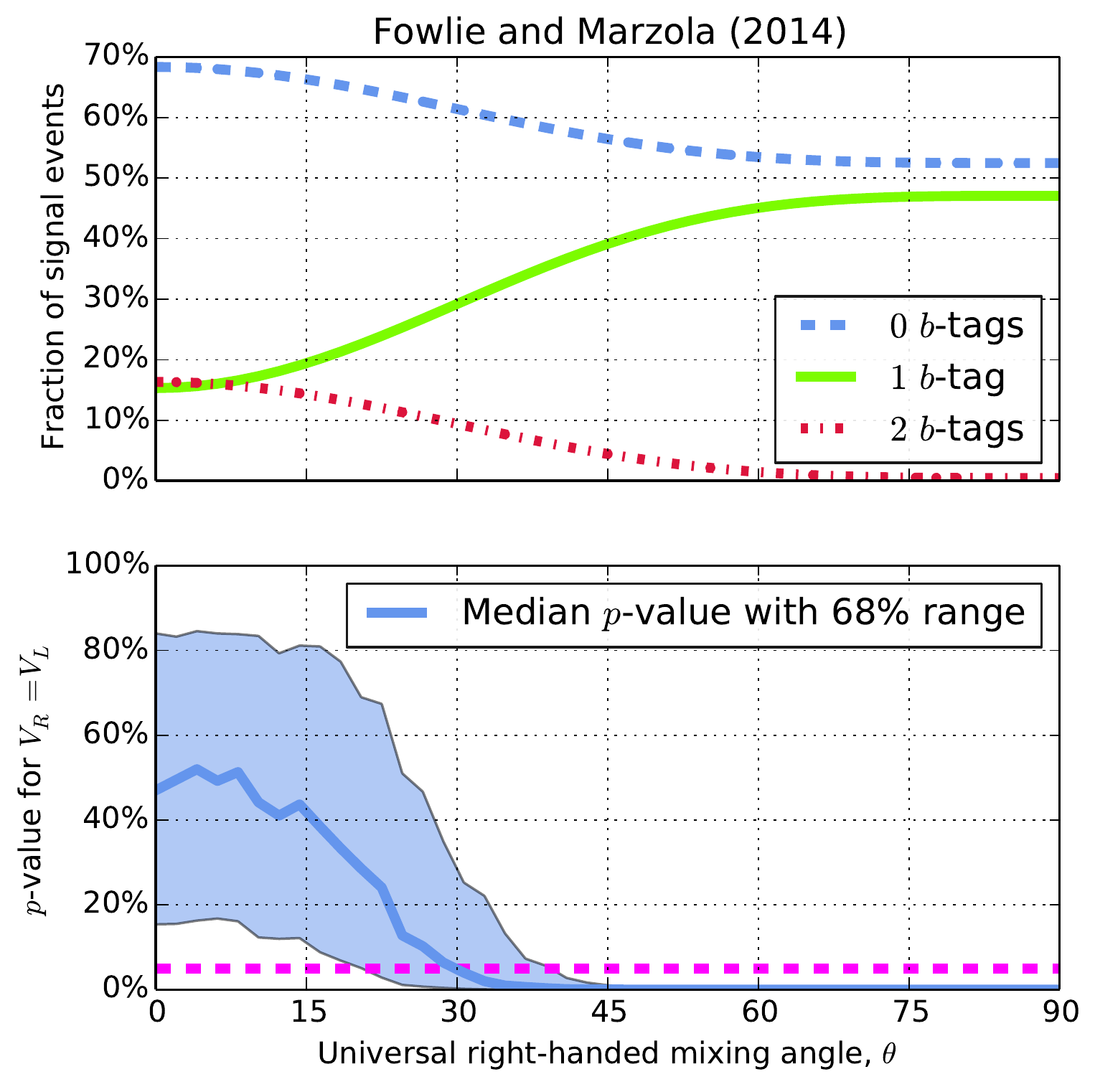}
}
\caption{A \roots{14} scenario with $\int\mathcal{L}\sim20\invfb$, with efficiency $\eff=0.7$ and purity $\purity=0.99$. \protect\subref{fig:2d} Exclusion of the null hypothesis that $V_L = V_R$ on the $(\theta_{23},\theta_{13})$ plane. The LH CKM matrix is marked with an arrow. \protect\subref{fig:1d} The RH mixing matrix universal mixing angle against (upper) the expected fractions of signal events in the $b$-tag categories and (lower) the median $p$-value for the null hypothesis that $V_L=V_R$. The blue band is the $68\%$ interval for the $p$-value, over MC experiments. The pink dashed line indicates a $p$-value of $5\%$. If the $p$-value drops below $5\%$, we can reject the null hypothesis with at least $95\%$ confidence.
}
\label{fig:pval}
\end{figure}

%%%%%%%%%%%%%%%%%%%%%%%%%%%%%%%%%%%%%%%%%%%%%%%%%%%%%%%%%%%%%%%%%%%%%%%%%%%%%%%%
In \reffig{fig:2d}, we plot the median exclusion with $20\invfb$ at \roots{14} for the null hypothesis that the LH CKM matrix equals the RH mixing matrix, were the RH mixing matrix in fact described by independent $\theta_{13}$ and $\theta_{23}$ mixing angles.
In the white region in \reffig{fig:2d}, the $\theta_{13}$ and $\theta_{23}$ mixing angles yield a median $p$-value that is greater than $0.05$ ($2\sigma$); were the experiment repeated many times, in more than $50\%$ of circumstances the hypotheses $V_R = V_L$ and $V_R \neq V_L$ are distinguished with a confidence level equal or less than $2\sigma$. In other words, in more than $50\%$ of circumstances, the numbers of observed events across our $b$-tag categories would be statistically compatible with both the mentioned hypotheses. Thus, in this case, our test cannot find discrepancies between the quark mixing matrices in the two chiral sectors. The situation is different in the dashed and squared regions in \reffig{fig:2d}, in which the median $p$-value is less than $0.05$ (excluded at $2\sigma$) and $0.003$ (excluded at $3\sigma$), respectively: the actual mixing angles in $V_R$ are such that in more than $50\%$ of circumstances, the hypothesis $V_R=V_L$ can be rejected by at least $2\sigma$.
For instance, if either of the $\theta_{13}$ and $\theta_{23}$ RH mixing angles were greater than about $40\degree$ or if both were greater than about $30\degree$, we expect that in at least $50\%$ of circumstances our method is able to distinguish between the hypotheses $V_R = V_L$ and $V_R \neq V_L$. 
    
Because the $p$-value is invariant under the exchange $\theta_{13}\leftrightarrow\theta_{23}$, \reffig{fig:2d} is expected to be symmetric about the diagonal. In practice, \reffig{fig:2d} is approximately spherically symmetric so the quantitative behavior of the $p$-value can be illustrated by simply picking a direction on the $(\theta_{23}, \theta_{13})$ plane. Thus we choose a common value for the mixing angles, the \emph{universal mixing angle} $\theta \equiv \theta_{13} = \theta_{23}$, and plot in the upper panel of \reffig{fig:1d} the expected fraction of signal events in each $b$-tag category against this quantity. If the mixing angles in $V_R$ are identical and  below $15 \degree$, corresponding to $\theta \lesssim 15 \degree$, we expect that about 70\% of the signal events will carry 0 $b$-tags. The remaining 30\% will instead be evenly shared between the 1 $b$-tag and 2 $b$-tags categories. Conversely, when $\theta \gtrsim 75 \degree$, we expect that the $W_R$ hadronic decays yield a negligible amount of 2 $b$-tags events and about $50\%$ 0 $b$-tag and $50\%$ 1 $b$-tag events. 

The lower panel of \reffig{fig:1d} shows the median $p$-value (solid blue line) for the hypothesis $V_R = V_L$ as a function of the universal mixing angle. The light blue band represents the corresponding $68\%$ ($1 \sigma$) interval\footnote{Were the experiment repeated many times, in $68\%$ of experiments the observed $p$-value would fall in that interval.}  obtained in our MC simulations, while the magenta dashed line signals the $5\%$ (corresponding to 2$\sigma$) exclusion level. Whenever the median $p$-value drops below $5\%$, in at least $50\%$ of the circumstances the experiment will observe a discrepancy greater or equal to 2$\sigma$ against $V_R = V_L$. When the whole $1 \sigma$ band drops below $5\%$ the former conclusion applies to $84\%$ of the circumstances.      
For instance, our method reveals that if $\theta\gtrsim30\degree$ ($\theta\gtrsim40\degree$), in at least $50\%$ ($84\%$) of circumstances the null hypothesis $V_R = V_L$ will be rejected with at least $95\%$ --- \ie $2\sigma$ --- confidence. 

Thus, it appears that with limited integrated luminosity of about $20\invfb$ at \roots{14}, it might be possible to reject the theory that the LH CKM matrix is equal to the RH mixing matrix. Whether this is possible is, of course, dependent on the size of the mixing angles in the RH mixing matrix. If the $\theta_{13}$ and $\theta_{23}$ mixing angles in the RH mixing matrix differ only slightly from those in the LH CKM matrix, it will be difficult to test the equality of the LH CKM matrix and the RH mixing matrix with the considered luminosity. On the other hand, as explained in \refsec{sec:Method-LR}, nothing can be inferred about the mixing angle between light quarks, $\theta_{12}$, or complex phases.

%%%%%%%%%%%%%%%%%%%%%%%%%%%%%%%%%%%%%%%%%%%%%%%%%%%%%%%%%%%%%%%%%%%%%%%%%%%%%%%%
%%%%%%%%%%%%%%%%%%%%%%%%%%%%%%%%%%%%%%%%%%%%%%%%%%%%%%%%%%%%%%%%%%%%%%%%%%%%%%%%

To investigate the full potential of the LHC within the proposed framework, we consider two additional scenarios: a scenario with an increased integrated luminosity of $\int\mathcal{L}\sim3000\invfb$ and a scenario with $\int\mathcal{L}\sim3000\invfb$ and an improved in $b$-tagging efficiency $\eff=0.8$ to the detriment of the purity, $\purity=0.98$. 

\begin{figure}[t]
\centering
\subfloat[][$\int\mathcal{L}\sim3000\invfb$, with efficiency $\eff=0.7$ and purity $\purity=0.99$]{\label{fig:l}
\includegraphics[width=.48\textwidth]{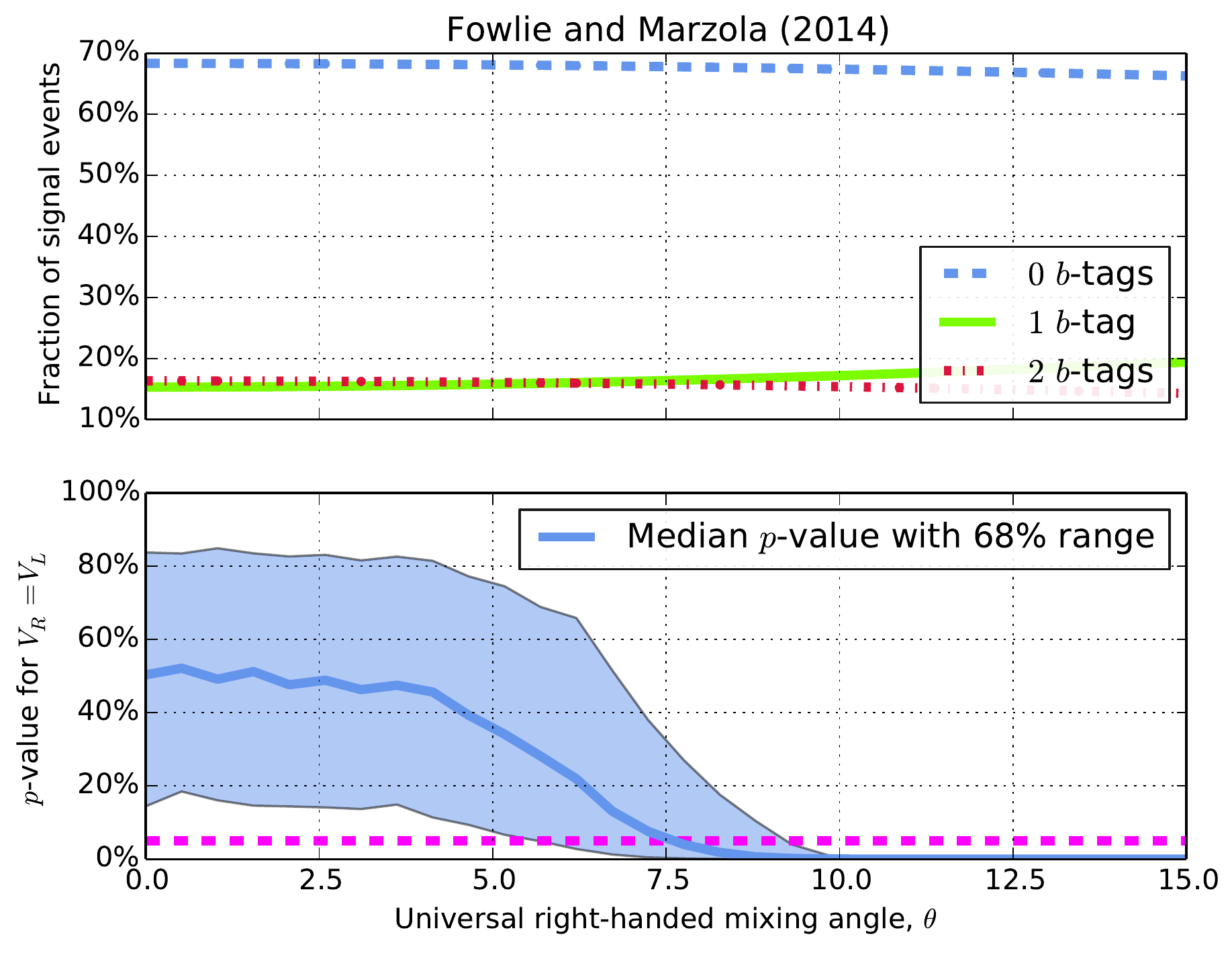}
}
\subfloat[][$\int\mathcal{L}\sim3000\invfb$, with efficiency $\eff=0.8$ and purity $\purity=0.98$]{\label{fig:r}
\includegraphics[width=.48\textwidth]{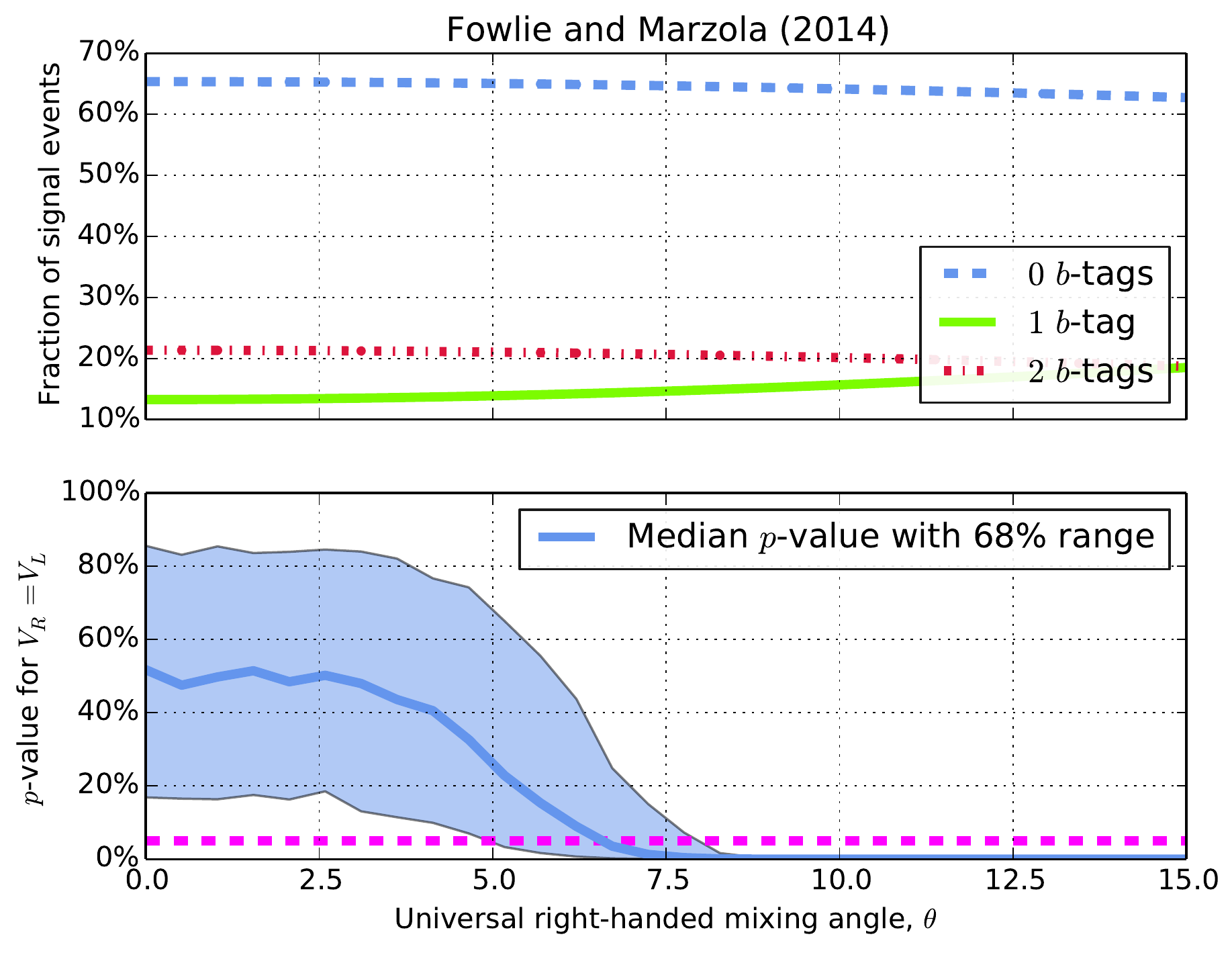}
}
\caption{As in \protect\reffig{fig:1d}, though for $\theta\leq15\degree$ and with $\int\mathcal{L}\sim3000\invfb$ (left) and with an improved efficiency (right). The RH mixing matrix universal mixing angle against (upper) the expected fractions of events in the $b$-tag categories and (lower) the median $p$-value for the null hypothesis that $V_L=V_R$. The blue band is the $68\%$ interval for the $p$-value, over MC experiments. The pink dashed line indicates a $p$-value of $5\%$. If the $p$-value drops below $5\%$, we can reject the null hypothesis with at least $95\%$ confidence.
}
\label{fig:inc}
\end{figure}

Then, in \reffig{fig:l}, we repeat the same analysis of \reffig{fig:1d} considering an improved integrated luminosity $\int\mathcal{L}\sim3000\invfb$. As clear from the bottom panel, the increased numbers of events result in sensitivity to a universal mixing angle as small as about $7.5\degree$. Below this threshold, as shown in the upper panel, the numbers of events in each $b$-tag category in the $V_R=V_L$ and $V_L\neq V_R$ hypotheses are too similar for the hypotheses to be discriminated. 

Considering an improved $b$-tagging efficiency on top of an increased integrated luminosity, in the bottom panel of \reffig{fig:r} we make slight inroads into $\theta\lesssim7.5\degree$. The improved efficiency results in sensitivity to a universal mixing angle as small as about $6.5\degree$. With current algorithms, the assumed $b$-tagging efficiency is unrealistic, nevertheless our study suggests that slight improvements in $b$-tagging efficiency, even to the detriment of purity, could improve sensitivity to the RH mixing matrix.

Let us remark upon the pros and cons of our method compared with complementary experiments in flavor physics\cite{Barenboim:1996nd,Blanke:2011ry}. As we have shown, by counting $b$-tags we are sensitive to a combination of only two mixing angles and none of the complex phases in the RH mixing matrix. Thus, the complete exploration of the RH mixing matrix indeed requires additional information from low-energy flavor physics experiments. If a minimal model is assumed, RH mixing angles can be extracted from the latter with a precision significantly higher than that which can be achieved by our method with $\int\mathcal{L}\sim3000\invfb$. In this case, provided a suitable integrated luminosity is collected, the proposed method constitutes a necessary independent cross-check. On the other hand, within non-minimal LR models where cancellations between flavor-changing neutral currents might limit the sensitivity of flavor observables, the proposed test still provides access into the structure of the RH mixing matrix.

%%%%%%%%%%%%%%%%%%%%%%%%%%%%%%%%%%%%%%%%%%%%%%%%%%%%%%%%%%%%%%%%%%%%%%%%%%%%%%%%
%%%%%%%%%%%%%%%%%%%%%%%%%%%%%%%%%%%%%%%%%%%%%%%%%%%%%%%%%%%%%%%%%%%%%%%%%%%%%%%%

%%%%%%%%%%%%%%%%%%%%%%%%%%%%%%%%%%%%%%%%%%%%%%%%%%%%%%%%%%%%%%%%%%%%%%%%%%%%%%%%
\section{Conclusions and outlook}
%%%%%%%%%%%%%%%%%%%%%%%%%%%%%%%%%%%%%%%%%%%%%%%%%%%%%%%%%%%%%%%%%%%%%%%%%%%%%%%%
In light of an experimental hint from the LHC, left-right symmetric models are attracting renewed interest. In minimal left-right symmetric models, the LH CKM matrix is approximately equal to the RH mixing matrix. We proposed an experimental test of this equality at the LHC at \roots{14}, in a scenario in which a right-handed $W$-boson with a mass of about $2\tev$ had been discovered, as suggested by the hint. 

Our test involved counting the numbers of $b$-tags resulting from the right-handed $W$-boson's hadronic decays. We found that at \roots{14} with a limited integrated luminosity of about $20\invfb$, minimal left-right symmetric models could be rejected at $95\%$ confidence, if the mixing angles in the RH mixing matrix were greater than about $30\degree$. Our test was, however, insensitive to complex phases and the mixing angle between the light quarks. With an increased integrated luminosity of about $3000\invfb$, our test was sensitive to RH mixing angles as small as about $7.5\degree$ and less if $b$-tagging efficiencies could be improved or optimized for our test. 

Because in this paper we simply proposed a method, we made conservative approximations in our analysis. In particular, rather than performing a full Monte-Carlo simulation of the test, we scaled background estimates from the quoted CMS study and modelled the former on the most dangerous background, $t \bar t$ in our case.   
In a forthcoming publication\cite{prepAL}, we will propose a similar experimental test of the unitarity of the RH mixing matrix. 

\begin{acknowledgements}
We thank E. Gabrielli, M. Heikinheimo, M. Raidal and C. Spethmann for the useful comments and discussions. 
AF was supported in part by grants IUT23-6, CERN+,  and by the European Union
through the European Regional Development Fund and by ERDF project 3.2.0304.11-0313
Estonian Scientific Computing Infrastructure (ETAIS).
LM acknowledges the European Social Fund for supporting his work with the grant MJD387.
\end{acknowledgements}

\bibliography{testing_VR}
\end{document}